\documentclass[14pt]{extarticle}
\usepackage{amsmath}
\usepackage{graphicx} 
\usepackage[margin=2.5cm]{geometry}
\usepackage{cite}
\DeclareUnicodeCharacter{2212}{-}

\title{The Mathematics of Chlorophyll Photoexcitation}
\author{Archit Chaturvedi
\\ University of California, Berkeley
\\ architc@berkeley.edu}
\date{March 2024}

\begin{document}

\maketitle
\begin{abstract}
    Photosynthesis is a fundamental process for plants to produce energy and survive. It is a well known fact that the light reactions in photosynthesis are a significant part of the overall process, and are carried out by chlorophyll molecules. This paper outlines the quantum chemical and mathematical description concerning electronic state transitions that take place in chlorophyll molecules during photosynthesis. In particular, the electronic transition of chlorophylls is assumed as a diatomic rovibrational $0 \longrightarrow 0$ transition, and is mathematically described accordingly. Then, the correlation of the Franck-Condon Principle in accordance to the photoexcitation of chlorophylls is delineated. Finally, existing research concerning the concepts laid out in this paper is discussed, and future implications of quantum chemical theory concerning chlorophyll photoexcitation is discussed. 
\end{abstract}
\section{Introduction}
Photosynthesis is a fundamental process in plants. It is responsible for producing chemical energy in plants to drive their survival. Photosynthesis is divided into two parts: 1. Light Reactions and 2. Calvin Cycle ([6]). In the context of this paper, light reactions are of focus. Chlorophylls are the most significant molecule when discussing light reactions in photosynthesis. Chlorophylls are located in the thylakoid membranes of a plant's chloroplast. These chlorophylls are photoexcited through absorbing photons, and it is this photoexcitation of chlorophylls that ultimately drives photosynthesis. Of the chlorophyll molecules, \textit{chlorophyll a} is the most significant molecule, as it is the key pigment in a plant's light reactions. There are other types of chlorophylls, such as chlorophyll b, which are discussed towards the end of the paper. Ultimately, theoretical formulations in quantum chemistry can be used to mathematically describe the photoexcitation of chlorophyll molecules. This paper aims to elucidate the mathematics behind such electronics transitions that take place in a chlorophyll molecule. One model is to assume the electronic transitions of chlorophylls as diatomic rovibrational transitions ([1, 2, 3]). This can give the total energy of a chloroplast's electronic state, and can also describe the transition energy required for the photoexcitation of chloroplasts. Furthermore, the Franck-Condon principle is significant when discussing electronic transitions ([1, 4, 5]). It describes electronic transitions under a fixed nuclear position, and can give the probability that an electronic transition occurs based on nuclear overlap. 
\section{Electronic Transitions in Chlorophylls}
We can view the electronic transition of a chlorophyll molecule as a diatomic rovibrational transition; that is, a transition with both rotations and vibrations. Approximating based on an anharmonic oscillator model with a nonrigid rotator, the energy of the chlorophyll's ground state can be written as: 
\begin{eqnarray}
    E_{ground} = G(g) + F(J) 
    \\ = \nu_{g}\left( g + \frac{1}{2} \right)-\chi_g\nu_g\left( g+\frac{1}{2} \right)^2 + B_{g}J_{g}(J_{g}+1) - D_{g}J_{g}^2(J_{g}+1)^2
\end{eqnarray}
In this equation, $\nu_{g}$ denotes the vibration frequency at ground state, $\chi_g$ denotes the anharmonic constant at ground state, $g$ denotes the vibrational quantum number at ground state, $B_g$ denotes the rotational constant at ground state, $J_{g}$ denotes the rotational quantum number of the ground state, and $D_{g}$ denotes the centrifugal distortion constant at the ground state. Ultimately, $G(g)$ marks the vibrational energy of the molecule, and $F(J)$ marks the rotational energy of the molecule. Denoting the total energy at the excited state requires simply adding the change in electronic transition energy, denoted as $\nu_{el}$:
\begin{equation}
    E_{excited} = \nu_{el} + \nu_{g}\left( g + \frac{1}{2} \right)-\chi_g\nu_g\left( g+\frac{1}{2} \right)^2 + B_{g}J_{g}(J_{g}+1) - D_{g}J_{g}^2(J_{g}+1)^2
\end{equation}
The eigenstate to eigenstate transition between the ground state and excited state can be written as: 
\begin{equation}
    E_{transition} = E_{excited} - E_{ground}
\end{equation}
Assuming a $0\longrightarrow 0$ transition between the ground state and excited state would be the most reasonable in the case of chlorophylls, as this is the most common and significant type of electronic transition. Ultimately, in such a case, there would be no vibrational change, and so, letting $\nu_{e}$ denote the vibrational frequency of the excited state and letting $\chi_{e}$ denote the anharmonic constant of the excited state, the total transition frequency can be written as: 
\begin{equation}
    E_{transition} = T_{el}+\left( \frac{1}{2}\nu_g - \frac{1}{4}\chi_g\nu_{g} \right) - \left( \frac{1}{2}\nu_{e} - \frac{1}{4}\chi_{e}\nu_{e}\right)
\end{equation}
In the equation above, it should be noted that $E_{transition}$ represents the transition frequency, or the overall difference in energy between the two electronic and vibration states. And $T_{el}$ represents the electronic transition energy. The rest of the equation marks the vibrational contribution to the transition as a result of anharmonicity. Ultimately, $E_{transition}$ denotes the smallest possible energy to observe an electronic transition between the ground state and excited state in a chlorophyll molecule. And the electronic transition energy can be found by subtracting the vibrational component of equation 5 from $E_{transition}$.

\section{Franck-Condon Principle and Chlorophylls}
It would be rather vacuous to discuss electronic transitions in chlorophylls without including the Franck-Condon Principle. As described, light reactions within a chlorophyll molecule involve the absorption of a photon, which is ultimately what causes the electronic transition. Franck-Condon Principle describes the intensities of vibronic transitions by assuming a fixed nuclear position during such transitions. The probability of a transition occurring from the ground state to the excited state can be given as:
\begin{equation}
    \left\langle \psi_{p, e}|\mu| \psi_{p, g} \right\rangle = \left\langle \psi_{nuc, e} ^*|\psi_{nuc, g} \right\rangle \left\langle \psi_{el, e}^*|\mu|\psi_{el, g} \right\rangle
\end{equation}
The first operator represents the nuclear overlap of the transition, and the second operator represents the electronic component of the transition. Ultimately, if the nuclear overlap is 0, then the electronic transition will not occur, despite the electronic component of the equation. A harmonic oscillator model is perhaps the simplest of models used to denote nuclear overlap in electronic transitions based on the Franck-Condon principle. In such a model, the wavefunction of the ground electronic state can be written as: 
\begin{equation}
    \left| \psi_g (Q) \right\rangle = \left| \left( \frac{\alpha}{\pi} \right)^\frac{1}{4} \textbf{exp} \left( \frac{-\alpha(Q-Q_g)^2 }{2} \right)\right\rangle
\end{equation}
In the equation above, $\alpha$ is a parameter concerning the vibrational motion of the chlorophyll molecule: 
\begin{equation}
    \alpha = \frac{\sqrt{mk}}{\hbar },
\end{equation}
with $m$ being the reduced mass of the system and $k$ denoting the spring constant. Furthermore, $Q$ denotes the position of the nuclei in the molecule, and $Q_g$ denotes the equilibrium bond length of the ground electronic state. Based on $\left| \psi_g (Q) \right\rangle$, the wavefunction of the excited electronic state can also be written as:
\begin{equation}
    \left| \psi_e (Q) \right\rangle = \left| \left( \frac{\alpha}{\pi} \right)^\frac{1}{4} \textbf{exp} \left( \frac{-\alpha(Q-Q_e)^2 }{2} \right)\right\rangle
\end{equation}
From this, the nuclear overlap integral can be written as:
\begin{equation}
    S_{ge} = \sqrt{\frac{\alpha}{\pi}}\int_{-\infty }^{\infty}\textbf{exp}\left( \frac{-\alpha\left( 2Q^2 -QQ_g-2QQ_e+Q_g^2 + Q_e^2\right)}{2} \right) dQ
\end{equation}
Ultimately, since:
\begin{equation}
    Q_g^2 + Q_e^2 = \frac{1}{2}\left[ (Q_g+Q_e)^2 + (Q_g-Q_e)^2 \right],
\end{equation}
the integral turns out to be a Gaussian integral, and so therefore, $S_{ge}$ can finally be expressed as:
\begin{equation}
    S_{ge} = \textbf{exp}\left( \frac{-\alpha(Q_g - Q_e)^2}{4} \right)
\end{equation}
This represents the value of the nuclear overlap of an electronic transition in a chlorophyll molecule and can be used to calculate the probability of chlorophyll photoexcitation given the electronic components of the chlorophyll molecule.
\section{Existing Research and Future Implications}
There has been quite some research done on the quantum chemical properties of chlorophylls. The vibrational properties of chlorophylls have been explored by numerous studies. For example, source [7] calculated the vibrational frequencies for chlorophyll a-4, chlorophyll a4+, chlorophyll a5, and chlorophyll a5+. Furthermore, source [8] calculated the vibrational frequencies associated witht he excited state of chlorophyll f and chlorophyll a based on the Franck-Condon Principle. Sources [9] and [10] explore further chemical properties of chlorophyll a and bacteriochlorophyll a in accordance to the Franck-Condon principle. However, while much research has been done concerning the electronic transitions and quantum chemical properties of chlorophyll molecules, further research is needed to be done to fully understand the overall picture of photoexcitation in chlorophylls during photosynthesis. For example, using the models discussed in this paper, while vibrational frequencies for chlorophylls have been calculated, the anharmonic constant of chlorophyll molecules still needs to be researched upon further. Also, the rotational component of electronic transitions in chlorophyll molecules needs further experimental research. Research concerning the equilibrium bond length of the ground/excited states of different types of chlorophyll molecules, as well as the orientation of atomic nuclei in different chlorophyll molecules may also be useful when discussing the photoexcitation of chlorophyll molecules. Finally, since there are so many different types of chlorophyll molecules out there, research concerning the electronic transitions in such chlorophylls, as well as the Franck-Condon components of these chlorophyll molecules, would open up exciting discoveries and opportunities concerning the photoexcitation of chlorophylls during light reactions.


\begin{thebibliography}{9}

\bibitem{Tuckerman}
Mark Tuckerman, Quantum Chemistry, 2024

\bibitem{Tully}
Tully, John C. "Molecular dynamics with electronic transitions." The Journal of Chemical Physics 93.2 (1990): 1061-1071.

\bibitem{Kasha}
Kasha, Michael. "Characterization of electronic transitions in complex molecules." Discussions of the Faraday society 9 (1950): 14-19.

\bibitem{Coolidge}
Coolidge, Albert Sprague, Hubert M. James, and Richard D. Present. "A study of the Franck‐Condon principle." The Journal of Chemical Physics 4.3 (1936): 193-211.

\bibitem{Condon}
Condon, Edward U. "The Franck-Condon principle and related topics." American journal of physics 15.5 (1947): 365-374.

\bibitem{Campbell}
Reece, Jane B., Lisa A. Urry, and Michael L. Cain. Campbell biology. Pearson, 2017.

\bibitem{Parameswaran}
Parameswaran, Sreeja, Ruili Wang, and Gary Hastings. "Calculation of the vibrational properties of chlorophyll a in solution." The Journal of Physical Chemistry B 112.44 (2008): 14056-14062.

\bibitem{Zamzam}
Zamzam, Noura, and Jasper J. van Thor. "Excited state frequencies of chlorophyll f and chlorophyll a and evaluation of displacement through franck-condon progression calculations." Molecules 24.7 (2019): 1326.

\bibitem{Zazubovich}
Zazubovich, V., I. Tibe, and G. J. Small. "Bacteriochlorophyll a Franck− Condon Factors for the S0→ S1 (Q y) Transition." The Journal of Physical Chemistry B 105.49 (2001): 12410-12417.

\bibitem{Pieper}
Pieper, J., J. Voigt, and G. J. Small. "Chlorophyll a Franck− Condon factors and excitation energy transfer." The Journal of Physical Chemistry B 103.13 (1999): 2319-2322.


\end{thebibliography}
\end{document}